\documentclass[aps,prl,floatfix,preprintnumbers,twocolumn,superscriptaddress,nofootinbib]{revtex4-1}
\usepackage{amsmath,amsthm,amssymb,color,psfrag,url,latexsym,graphicx,epstopdf,slashed,xspace,hyperref,enumitem}
\usepackage{lineno}

\definecolor{darkred}{rgb}{0.6,0.0,0.0}
\definecolor{darkblue}{rgb}{0.0,0.0,0.5}
\definecolor{darkgreen}{rgb}{0.0,0.5,0.0}
\definecolor{brown}{rgb}{0.0,0.0,0.0}

\newcommand{\be}{\begin{equation}}
\newcommand{\ee}{\end{equation}}
\newcommand{\bea}{\begin{eqnarray}}
\newcommand{\eea}{\end{eqnarray}}

\begin{document}
\preprint{MIT-CTP 4973}
\title{Factorization of Standard Model cross sections at ultra high energy}

\author{Yang-Ting Chien}
\email{ytchien@mit.edu}
\affiliation{
Center for Theoretical Physics, Massachusetts Institute of Technology,
Cambridge, MA 02139, USA}
\author{Hsiang-nan Li}
\email{hnli@phys.sinica.edu.tw}
\affiliation{Institute of Physics, Academia Sinica,
Taipei, Taiwan 115, Republic of China}

\date{\today}

\begin{abstract}
The factorization theorem for organizing multiple electroweak boson
emissions at future colliders with energy far above the electroweak scale
is formulated. Taking the inclusive muon-pair production in
electron-positron collisions as an example, we argue that the
summation over isospins is demanded for constructing the universal
distributions of leptons and gauge bosons in an electron.
These parton distributions are shown to have the same infrared structure
in the phases of broken and unbroken electroweak symmetry,
an observation consistent with the Goldstone Equivalence Theorem.
The electroweak factorization of processes involving protons is
sketched, with an emphasis on the subtlety of the scalar distributions.
\end{abstract}
\maketitle

The discovery of the Higgs boson is one of the recent greatest achievements
in particle physics. The precise determination of Higgs boson properties,
the deeper understanding of the electroweak (EW) symmetry breaking mechanism,
and the continuous search for physics beyond the Standard Model (SM) then
motivate the construction of next-generation colliders, such as the Super
Proton-Proton Collider \cite{Arkani-Hamed:2015vfh,Tang:2015qga} and the
Future Circular Collider \cite{Mangano:2017tke}. The SM dynamics at energy
$E$ far above the EW scale, i.e., the vacuum expectation value (VEV)
$v$ of the Higgs field, produces new phenomena
\cite{Hook:2014rka} and impacts heavy dark matter searches
\cite{Ciafaloni:2010ti, Baumgart:2014vma,Bauer:2014ula,Ovanesyan:2014fwa}.
Multiple emissions of EW gauge bosons enhanced by the large radiative logarithms
$\ln (v/E)$ imply the invalidity of fixed-order
calculations \cite{Ciafaloni:1998xg,Chiu:2007yn}. It demands the treatment
of EW gauge bosons as partons, which inevitably alters the conventional
framework for SM studies. In this Letter we aim at constructing
a factorization theorem that incorporates the EW shower effects in SM processes
at ultra high energy.

The EW sector shares features similar to those of QCD but also exhibits
qualitative differences. The dynamically broken chiral symmetry in QCD gives
rise to hadron masses, while the spontaneously broken EW symmetry
induces particle masses. The VEV $v$
plays a role similar to the QCD scale $\Lambda_{\rm QCD}$, so that $v/E$
defines a power expansion parameter in the analysis of infrared EW
radiation. The isospin and hypercharge in the EW sector correspond to the
color in QCD. Due to the confinement, only color-singlet bound states can
be prepared and detected in experiments. We need to sum over colors
when connecting parton- and hadron-level formalisms. However, we can
distinguish particle flavors and thus encounter the
violation of the Block-Nordsieck theorem \cite{PhysRev.52.54,Ciafaloni:2000df}
and the aforementioned logarithms $\ln (v/E)$.

The QCD factorization has been intensively investigated for decades and
is a well-developed framework. The infrared EW logarithms
$\ln (v/E)$ can be handled in a way similar
to that of the QCD logarithms $\ln(\Lambda_{\rm QCD}/E)$:
the collinear logarithms are factorized into parton distribution functions
(PDFs), and the soft logarithms cancel to ensure the
universality of the PDFs. This soft cancellation is guaranteed by the
Kinoshita-Lee-Nauenberg theorem \cite{1962JMP.....3..650K,1964PhRv..133.1549L}
for ``inclusive-enough" cross sections. On the other hand, in the high energy
limit the mass term in the Higgs potential becomes irrelevant
with the approximate restoration of the EW symmetry. It is also possible
that a new ultraviolet theory with the exact
EW symmetry exists, which generates the SM Higgs potential radiatively
at low energy \cite{Cacciapaglia:2005da}. In either case, we can consider
an EW symmetry breaking scale $\mu_s$, above which the EW symmetry is
restored, and EW corrections contain infrared divergences.
Such restoration does not exist for the chiral symmetry
in zero-temperature QCD. With the additional scale $\mu_s$, as well as
scalar emissions through the Yukawa couplings in a huge hierarchy, the
power counting for the EW factorization is expected to be more
complicated than for the QCD one.

\begin{figure*}
    \includegraphics[width=0.3\textwidth]{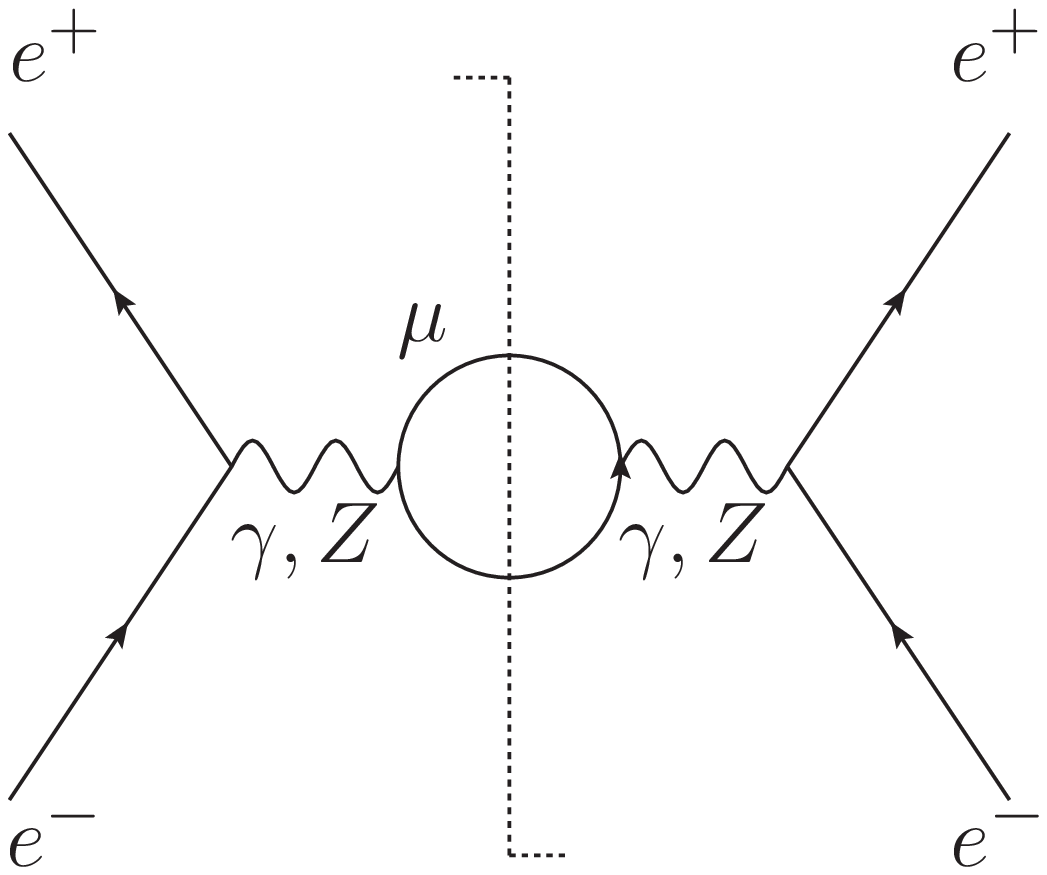}~
    \includegraphics[width=0.31\textwidth]{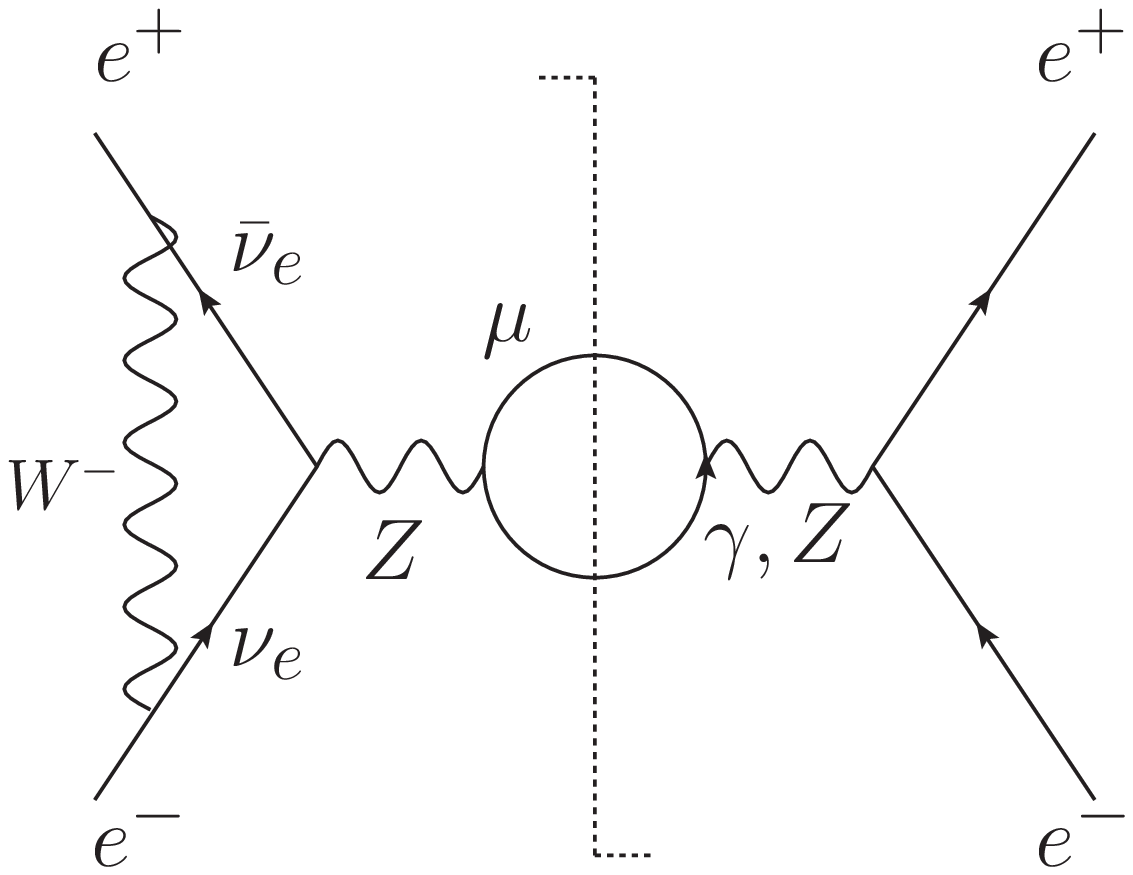}~
    \includegraphics[width=0.3\textwidth]{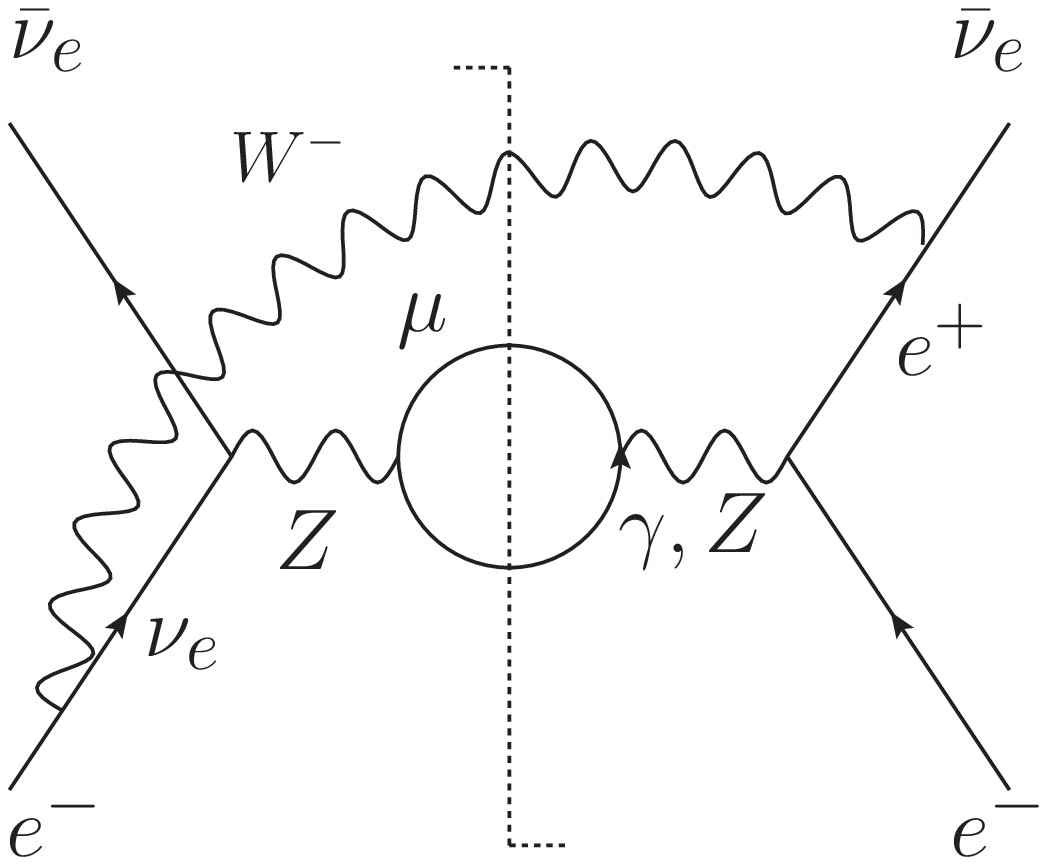}
    \caption{The LO diagram of the process
    $e^+e^- \to \mu^+\mu^-+X$ (left panel) and the NLO
    corrections with the $W^-$ boson exchange (middle and
    right panels) at the parton level. The dashed lines represent the final state cuts.}
\label{EW_LO}
\end{figure*}
\begin{figure*}
    \includegraphics[width=0.30\textwidth]{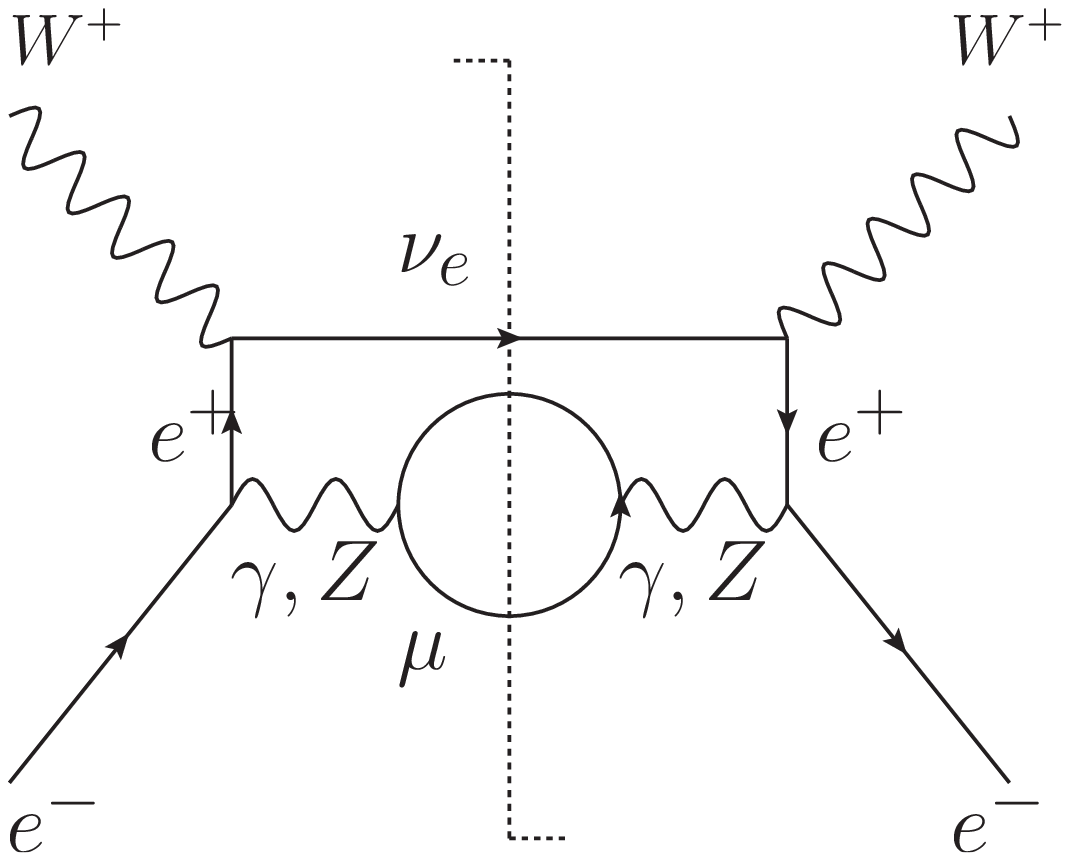}~
    \includegraphics[width=0.31\textwidth]{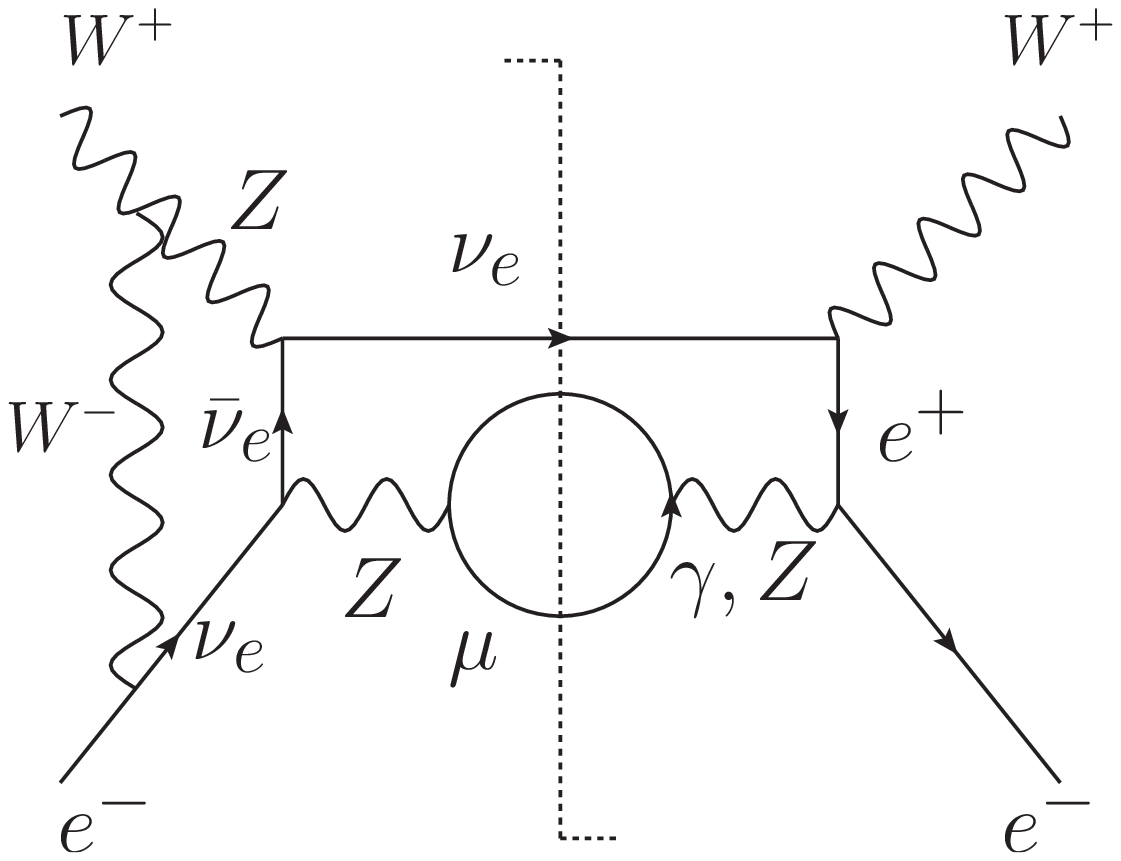}~
    \includegraphics[width=0.30\textwidth]{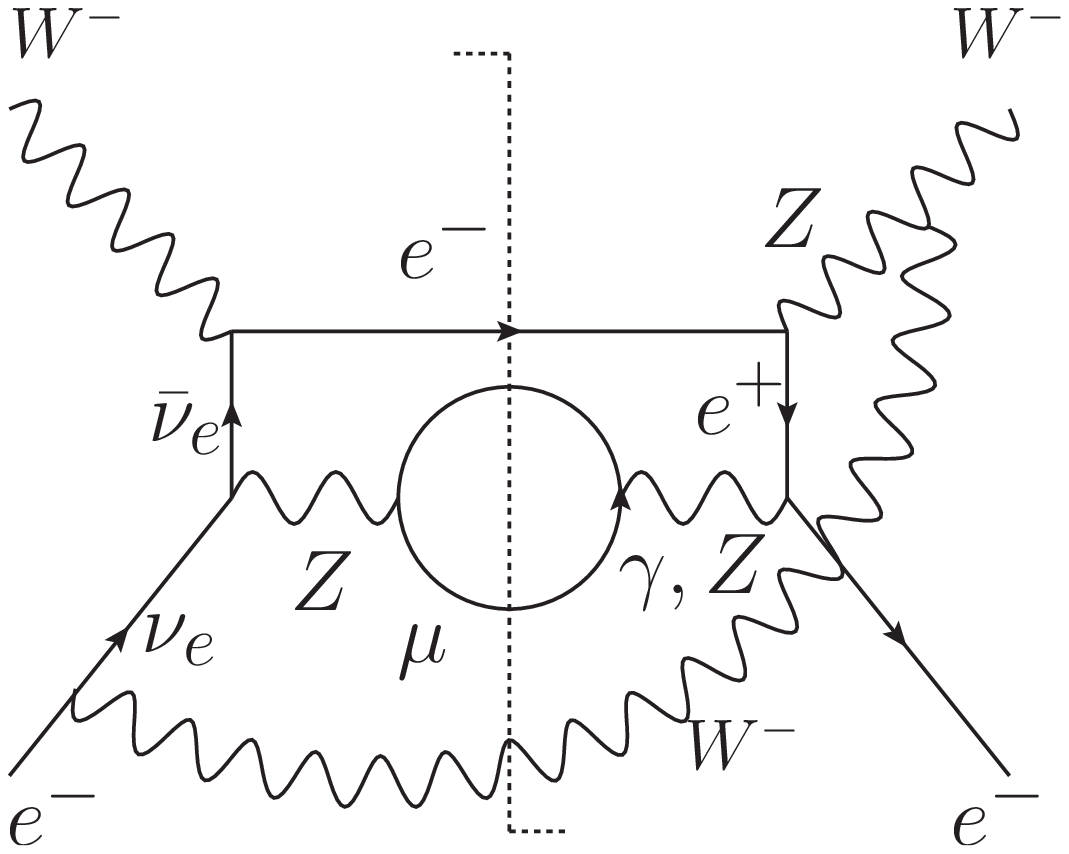}
    \caption{The LO diagram
    involving the parton $W^\pm$ of $e^+$ (left panel), and its NLO
    corrections with the $W^-$ boson exchange (middle and right panels).}
\label{EW_eW}
\end{figure*}

We analyze the EW infrared structure of
the inclusive production of muon pairs in electron-positron collisions
$e^- e^+ \to \mu^-\mu^+ + X$, and examine how EW emissions modify
the PDF definitions of an electron. It is shown that a
well-defined PDF at a scale $E\gg v$ requires the summation over the
$SU(2)_L\times U(1)_Y$ charges, i.e., isospins of partons. Adding EW
shower directly into the PDFs \cite{Chen:2016wkt,Bauer:2017isx} originally
designed to absorb QCD nonperturbative dynamics works only for $E$ not far
above $v$, where the EW radiation is not yet significant enough to violate
the universality. Furthermore, we argue that the PDFs have the same
infrared structure in the EW symmetry unbroken and broken phases, so it
suffices to implement a perturbative matching for the PDFs in the two
phases at the scale $\mu_s$. The EW factorization theorem then provides a
unified handle of the dynamics above and below $\mu_s$ up to power
corrections in $v/E$. It also allows us to resum the large logarithms
$\ln (v/E)$, a subject to be explored in the future.

For simplicity, we focus on the first lepton generation and turn off the
EW emissions from the muons. The construction of fragmentation
functions \cite{Baumgart:2014vma,Cavasonza:2014xra} follows the same reasoning.
The next-to-leading order (NLO) corrections \cite{Denner:2000jv,Denner:2001gw}
to the leading-order (LO) process
in the broken phase are displayed in Fig.~\ref{EW_LO}. The universality of
the PDFs holds only if the soft logarithms from gauge boson exchanges,
which couple the initial electron and positron beams, cancel between the
virtual and real corrections. The soft cancellation is straightforward in
the case, where a photon or a $Z$ boson attaches to the incoming partons
$e^-$ and $e^+$. In the virtual correction with the $W^-$ boson exchange, the
parton $e^-$ on the left side of the final state cut emits $W^-$, becoming a
neutrino $\nu_e$, and the parton $e^+$ absorbs $W^-$, becoming an
anti-neutrino $\bar \nu_e$. For the real correction, the other incoming
parton must be $\bar\nu_e$ produced through EW shower of $e^+$ in order to
annihilate $\nu_e$ into a $Z$ boson. To cancel the soft logarithms, the above
two NLO diagrams with different isospins $e^+$ and $\bar\nu_e$ have to be
added, indicating that we can define only the {\sl lepton distribution} of
the positron, $\phi_{\ell/e}$. A positron and an anti-neutrino can
frequently convert to each other by emitting $W$ bosons and form a lepton
beam before entering the hard process, to which both components of the lepton
doublet contribute.
\newline \indent As in the QCD factorization, collinear gauge bosons are
factorized by eikonalizing the parton lines the gauge bosons attach to, giving
the Wilson links which appear in the definition of the lepton PDF as a matrix
element of a nonlocal operator. Wilson links are crucial for the gauge invariance
of a PDF. Certainly, the diagrams with collinear gauge bosons attaching to hard
particle lines have to be included to complete the factorization.
Scalar emissions from the initial lepton in two-particle irreducible
diagrams (like those in Fig.~\ref{EW_LO}) are free of leading-power collinear
divergences, since the product of two adjacent
on-shell fermion propagators suppresses the collinear region.
It means that collinear scalars do not contribute to Wilson
links, namely, are irrelevant to the gauge invariance. 
Scalar emissions in two-particle reducible
diagrams, such as the self-energy correction of the lepton, can generate
leading-power infrared logarithms,
which are, however, suppressed by the Yukawa coupling in the present study.
\begin{figure*}
    \includegraphics[width=0.24\textwidth]{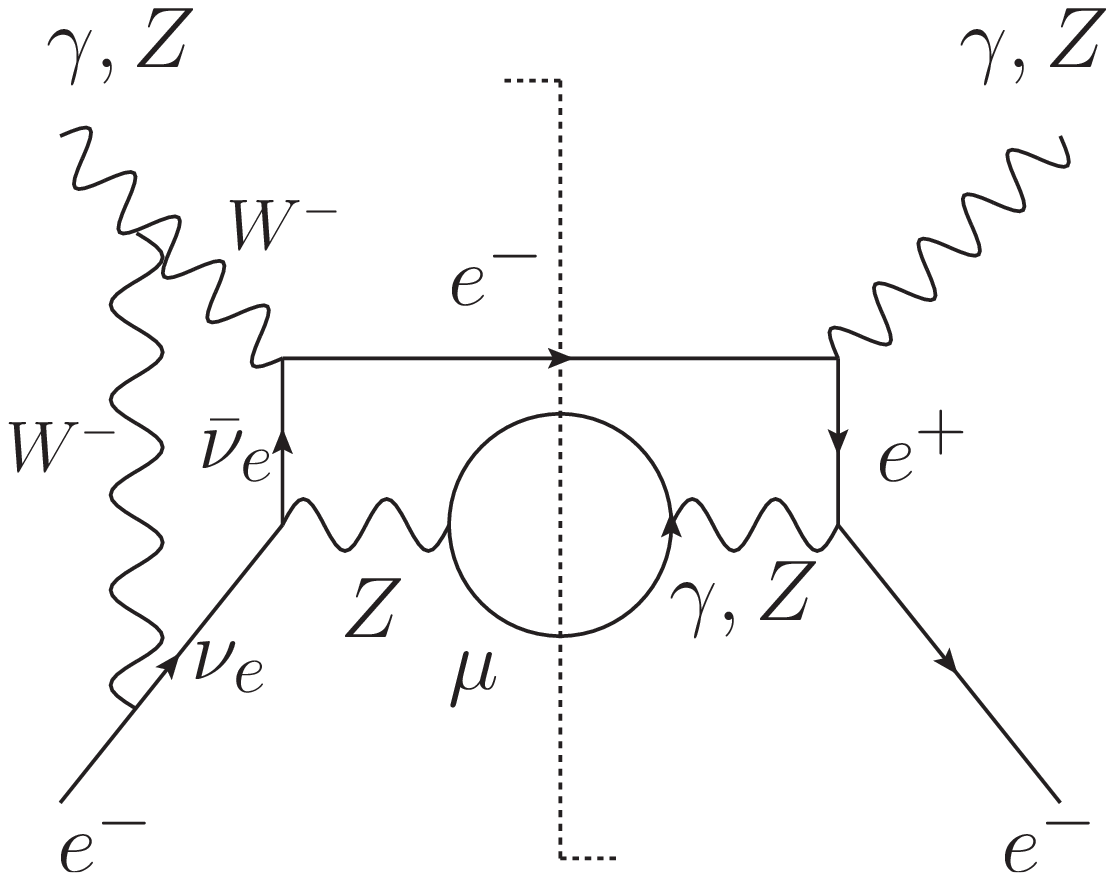}~
    \includegraphics[width=0.24\textwidth]{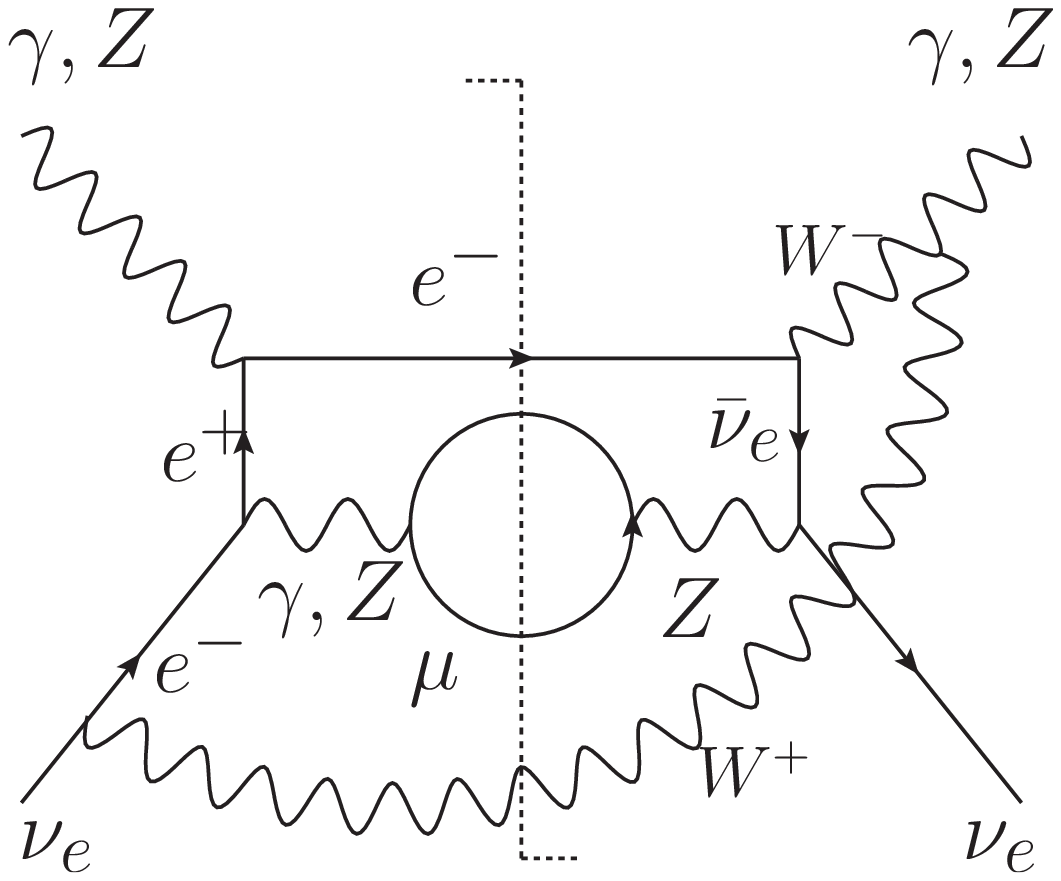}~
    \includegraphics[width=0.24\textwidth]{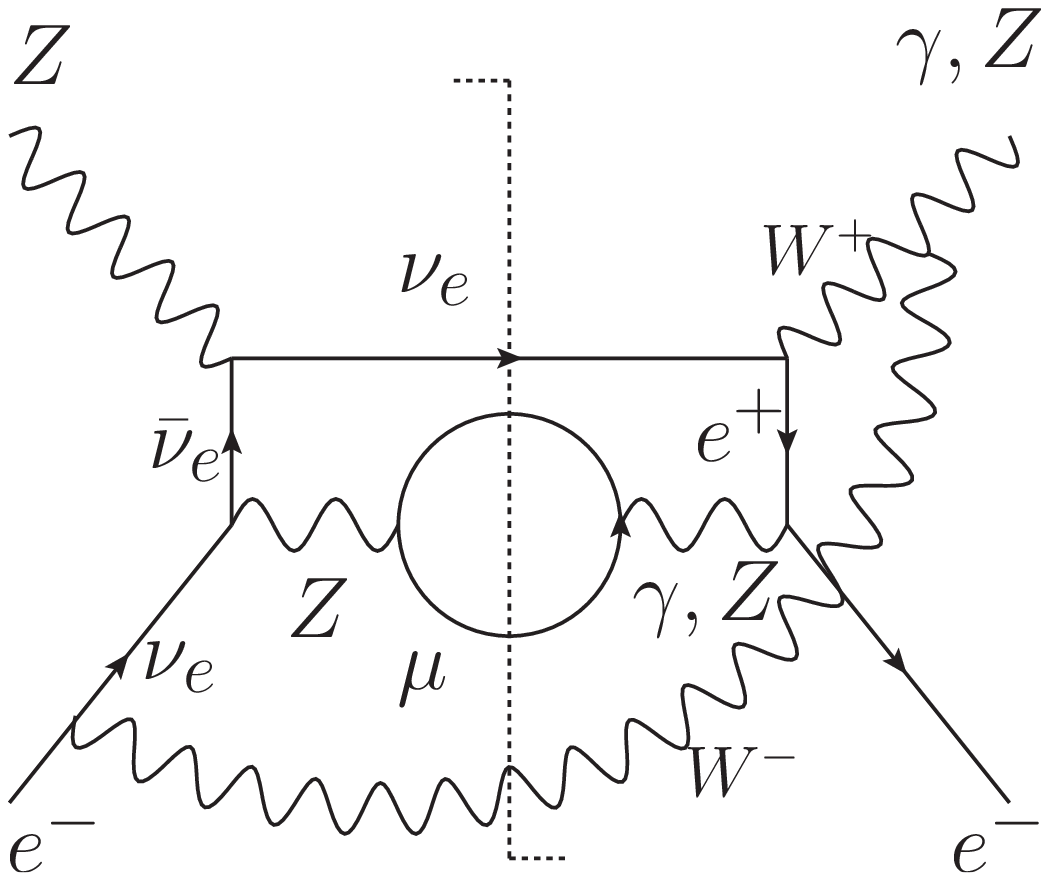}~
    \includegraphics[width=0.24\textwidth]{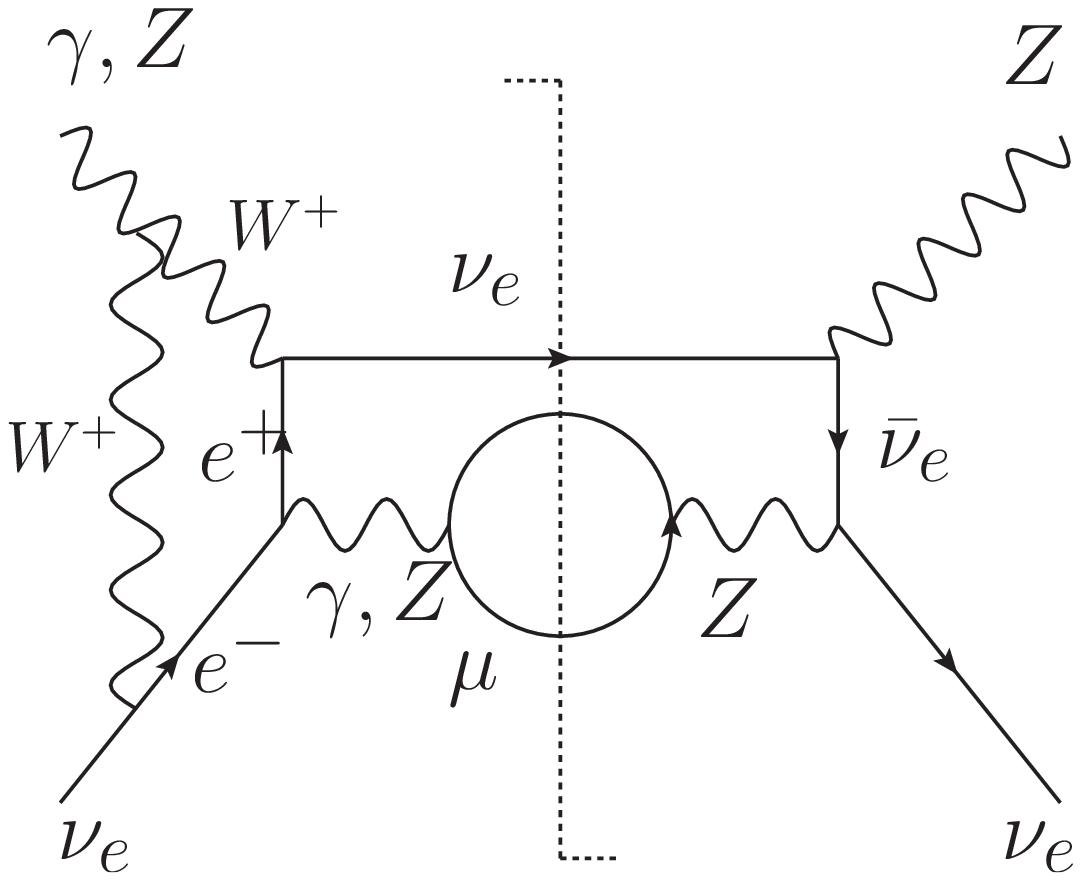}
    \caption{The NLO diagrams involving the parton $\gamma$ or $Z$ of $e^+$, with the $W$ boson exchange.}
\label{EW_NLO_gZ}
\end{figure*}
\newline\indent For the construction of gauge boson PDFs, the parton $e^+$
is replaced by a transversely polarized $W^+$ boson from EW
emissions. The contribution from a longitudinally polarized $W^+$ boson
is power-suppressed. The corresponding LO diagram is shown in the left
panel of Fig.~\ref{EW_eW}, where the flavor is labeled explicitly for each line.
At NLO, the exchange of a photon or a $Z$ boson has straightforward soft cancelation.
The exchanges of a virtual $W^-$ boson and a real $W^-$ boson are exhibited
in the middle and right panels of Fig.~\ref{EW_eW}, respectively.
The soft cancelation demands the summation of the two NLO diagrams
and the construction of the $W^\pm$ boson distribution, $\phi_{W/e}$.
Scalar emissions by a $W$ boson are power-suppressed by $v/E$
compared to gauge boson emissions, so Wilson links associated
with the $W$ boson PDF do not collect collinear scalars either.
\newline \indent The above soft cancelation does not involve the photon and $Z$
boson as incoming partons, hinting that the photon and $Z$ boson distributions
can be independent of the $W^\pm$ distribution. The construction of the photon
and $Z$ boson distributions is more involved. The incoming parton can be either
a photon or a $Z$ boson on each side of the final state cut,
so we need to consider the $\gamma$-$\gamma$, $\gamma$-$Z$,
$Z$-$\gamma$ and $Z$-$Z$ combinations. The virtual $W^-$ corrections in
Fig.~\ref{EW_NLO_gZ} contribute to all the combinations, while the real $W^-$
corrections contribute only to the $Z$-$\gamma$ and $Z$-$Z$ combinations. To have
the soft cancelation, we must include the partner parton $\nu_e$ of $e^-$.
The soft logarithms
in the virtual $W^-$ correction to the $e^-$-initiated process then cancel those
in the real $W^+$ correction to the $\nu_e$-initiated process. Similarly,
the soft logarithms in the real $W^-$ correction to the $Z$-$\gamma$ and $Z$-$Z$
combinations in the $e^-$-initiated process cancel those in the virtual
$W^+$ correction to the $\gamma$-$Z$ and $Z$-$Z$ combinations in the
$\nu_e$-initiated process.
For the $Z$ boson exchanges at NLO, the virtual $Z$
contributes to the $Z$-$\gamma$ and $Z$-$Z$ combinations, while the
real $Z$ contributes to the $\gamma$-$Z$ and $Z$-$Z$
combinations, among which the soft logarithms also cancel.
The above reasoning leads to the definitions of the photon, the $Z$ boson
and the mixed PDFs, $\phi_{\gamma/e}$, $\phi_{Z/e}$ and $\phi_{\gamma Z/e}$,
respectively.

The lepton distribution $\phi_{\ell/e}(x,\mu)$ of the electron
is written as
\begin{eqnarray}
& &\phi_{\ell/e}(x,\mu)=\frac{1}{4}\sum_{s}\int\frac{dy^-}{2\pi}\exp(-ixp^+y^-)\times\label{ell}\\
& &\hspace{0.5cm}\langle e(p,s)|\sum_{\ell=e,\nu_e}\bar\ell(y^-)W^\dagger(y^-)
\frac{1}{2}\gamma^+W(0)\ell(0)| e(p,s)\rangle,\nonumber
\end{eqnarray}
where $x$ is the momentum fraction, $\mu$ is the factorization scale,
and the Wilson link $W(y)$ is defined as
\begin{eqnarray}
W(y)&=&P\Big(\exp\left[ig\int_0^\infty dz n\cdot W_i(y+zn)\sigma_i\right]\times\nonumber\\
&&\exp\left[ig'\int_0^\infty dz n\cdot B(y+zn)I\right]\Big),
\end{eqnarray}
with the Pauli matrices $\sigma_i$, the identity matrix $I$,
the $SU(2)_L \times U(1)_Y$ gauge fields $W_i~(i=1,2,3)$ and $B$, and
the gauge couplings $g$ and $g'$ of $SU(2)_L$ and $U(1)_Y$, respectively.
The gauge eigenstates and the mass eigenstates can be transformed into each other,
$W_{1,2}\leftrightarrow W^\pm$ and $(W_3, B)\leftrightarrow (Z,\gamma)$,
but the expression will be more complicated if written in terms of the
latter due to the distinctive couplings.
It has been found that the difference between the up- and down-type
fermion distributions is driven to diminishing by a double-logarithmic EW evolution
\cite{Bauer:2017isx}. This result supports that we need to define only Eq.~(\ref{ell}),
whose associated hard function sums the contributions from both isospins.
The $W$ boson distribution is given by
\begin{eqnarray}
& &\phi_{W/e}(x,\mu)=\frac{1}{2}\sum_{s}\int\frac{dy^-}{2\pi xp^+}\exp(-ixp^+y^-)\times\\
&&\hspace{0.5cm}\langle e(p,s)|\sum_{i=1,2}W_{i\;\nu}^{+}(y^-)W^\dagger(y^-)
W(0)W^{\nu+}_{i}(0)| e(p,s)\rangle.\nonumber
\end{eqnarray}
The definitions for the photon, the $Z$ boson and
the mixed distributions are similar, with the field tensor $W_i^{\mu\nu}$
replaced by the corresponding ones.

When the factorization scale runs above the symmetry breaking
scale $\mu_s$, all particles become massless and some scalar degrees of
freedom emerge. It is then a concern how the infrared structure
of the process and the PDFs are modified.
Below we argue that the PDFs in the symmetry unbroken phase contain the same
collinear logarithms as in the symmetry broken phase.
For a massive gauge boson with momentum $k$ in the broken phase, we adopt
the propagator
\begin{eqnarray}
\frac{-i}{k^2-m_W^2}\left[g^{\mu\nu} -
\left(1-\frac{1}{\lambda}\right)\frac{k^\mu k^\nu}{k^2-m_W^2/\lambda}\right],
\end{eqnarray}
in the covariant gauge
with the gauge parameter $\lambda$. For a massless gauge boson in the
unbroken phase, we use
\begin{eqnarray}
\frac{-i}{k^2-m_W^2}\left[g^{\mu\nu} -
\left(1-\frac{1}{\lambda}\right)\frac{k^\mu k^\nu}{k^2}\right],
\end{eqnarray}
where the physical $W$ boson mass $m_W$ serves as an infrared regulator.
The PDFs, being gauge invariant, can be computed in an arbitrary gauge,
and we choose the Feynman gauge $\lambda\to 1$. The metric tensors
for real gauge boson emissions are given by
\begin{eqnarray}
g^{\mu\nu}-\frac{k^\mu k^\nu}{m_W^2},\;\;~~~
g^{\mu\nu}-\frac{k^\mu \bar k^\nu+k^\nu \bar k^\mu}{k\cdot \bar k},\label{metric}
\end{eqnarray}
in the broken and unbroken phases, respectively, with
$k=(k^0,\vec k)$ and $\bar k=(k^0,-\vec k)$. 
There exists cancelation with the corresponding virtual diagrams in the soft region.
The momentum $k$ picks up the minus lightcone component of the
contracted momentum, which is power suppressed in the collinear region.
Therefore, the second terms in Eq.~(\ref{metric}) can be dropped, when we focus on the infrared
logarithms. The Feynman rules for the gauge bosons are then identical in the
two phases in the collinear region with $k^2\to m_W^2$. Ghost propagators
are also identical in the covariant gauge. Leptons remain massless in
both phases at leading power. The above argument applies to the emissions of
photons and $Z$ bosons. In the end, the Feynman rules for cross section calculations in both
phases can be made identical in the infrared region, implying that
the infrared structure of the PDFs is not affected by the symmetry
breaking at the scale $\mu_s\gg v$. That is, the matching of the PDFs at $\mu_s$
is perturbative, a key observation for the EW factorization in the high energy limit.
\newline \indent We elaborate on the connection between longitudinally polarized
$W$ bosons in the broken phase and charged scalars in the unbroken phase.
The longitudinal polarization vector of the former
can be well approximated by $k^\mu/m_W$ at high energy below $\mu_s$.
When this vector is contracted with the lepton-gauge-boson vertex,
the Ward identity yields the suppression by powers of $m_\ell/v$,
$m_\ell$ being a lepton mass.
In the unbroken phase, scalar emissions from the lepton are suppressed by
the tiny Yukawa coupling, equivalent to
$m_\ell/v$ in the broken phase. This is a consequence of the Goldstone
Equivalence Theorem \cite{PhysRevD.16.1519}. Corrections to the Goldstone
Equivalence Theorem observed in \cite{Chen:2016wkt} are actually of higher power,
even though they are numerically important around the EW scale. Note that
collinear gauge boson emissions from scalars are of leading power, while collinear
scalar emissions from gauge bosons and fermions are power suppressed.
The scalar PDFs remain of higher power under
the evolution with the factorization scale, which involves either
higher power initial conditions or splitting kernels.
\newline \indent The factorization formula for $e^- e^+ \to \mu^-\mu^+ + X$ takes the form
\begin{eqnarray}
    \frac{d\sigma^{\mu^+\mu^-}}{dp_Tdy}&=&\sum_{i,j=\ell,b}\int dx_idx_j 
    \phi_{i/e^+}(x_i,\mu)\phi_{j/e^-}(x_j,\mu)\times\nonumber\\
     &&H_{i,j\rightarrow \mu^+\mu^-+X}(x_i,x_j,\mu)+{\cal O}(v/E)\;,\label{fact}
\end{eqnarray}
where
$b=W, \gamma, Z, \gamma Z$, and the hard function
$H_{i,j\rightarrow \mu^+\mu^-+X}$ describes the muon-pair production
with the transverse momentum $p_T$ and the rapidity $y$
at the parton level.
The double logarithms are absent in the EW evolution of $\phi_{\ell/e}$ as a
consequence of the soft cancellation illustrated before, once
the up- and down-type fermion distributions in \cite{Bauer:2017isx} are added. 
The standard PDFs from the EW evolution \cite{Ciafaloni:2005fm,Chen:2016wkt,Bauer:2017isx}
to a high scale $\mu > v$ can serve as the inputs of our PDFs in Eq.~(\ref{fact})
according to the isospin sum.
The PDFs at the parton level up to one loop can be obtained via
the LO EW splitting functions, for example,
\begin{eqnarray}
\phi_{\ell/\ell}(x,\mu)&=&\delta(1-x)
+\Big(\frac{3}{4}g^2+g'^2Y^2\Big)\frac{1}{8\pi^2}\ln\Big(\frac{\mu^2}{v^2}\Big)\times\nonumber\\
&&\Big[\frac{1+{x}^2}{(1-x)_+}+\frac{3}{2}\delta(1-x)\Big]\;,
\end{eqnarray}
from $\ell \to \ell+b$, in which $Y$ is the hypercharge, and
the collinear logarithmic pieces are the same in both the broken and unbroken phases. 
\newline \indent To conclude, we have explored the infrared structure of EW emissions in
electron-positron collisions in the high energy limit, and constructed
the universal PDFs which facilitate the EW factorization.
At ultra high energy, the significant EW shower makes initial state
beams of isospin doublets, such that
distinguishing the up- and down-type flavor distributions becomes
unpractical. Though an electron is not a $SU(2)$ singlet, the electron beam
as the result of the EW evolution to a high scale allows the definition
of its PDFs in terms of the isospin sum and the cancelation of soft EW emissions.
We have pointed out that these PDFs possess the same collinear
logarithms in the broken and unbroken phases, and that the perturbative
matching at the EW symmetry breaking scale is a new feature
compared to the QCD factorization.
For high-energy collisions involving protons, a quark PDF of the proton
should be defined with the summation over colors and isospins.
Besides, scalar emissions from the top quark parton can induce leading-power
infrared logarithms because of its large Yukawa coupling, implying that the
scalar PDFs of a proton need to be introduced.
Our formalism provides a theoretical framework for studying SM cross sections
far above the EW scale, and abundant extensions and applications are foreseen.

\section{Acknowledgements}
We thank C. Bauer, J.M. Chen, T. Han,
M. Mangano, E. Mereghetti and V. Vaidya
for helpful discussions. Y.-T. Chien thanks Los Alamos National Laboratory
for the support during which part of this work was performed. 
This work was supported in part by the Ministry
of Science and Technology of R.O.C. under Grant No. MOST-104-2112-M-001-037-MY3,
the US Department of Energy (DOE), Office of Science under Contract
No. DE-AC52-06NA25396, the DOE Early Career Program and the LHC Theory
Initiative Postdoctoral Fellowship under the National Science Foundation grant PHY-1419008.

\bibliography{ref}
\end{document}